\def\ts{\thinspace}

\def\simle{\thinspace\hbox{\raise.3ex\hbox{$<$} \llap{$_\sim$~}}\negthinspace}
\def\simge{\thinspace\hbox{\raise.3ex\hbox{$>$} \llap{$_\sim$~}}\negthinspace}

\newcommand{\ewiii}{\mbox{$EW(\lambda_{8498})$}}
\newcommand{\ewii}{\mbox{$EW(\lambda_{8542})$}}
\newcommand{\ewi}{\mbox{$EW(\lambda_{8662})$}}
\newcommand{\caiii}{\mbox{$\lambda_{8498}$}}
\newcommand{\caii}{\mbox{$\lambda_{8542}$}}
\newcommand{\cai}{\mbox{$\lambda_{8662}$}}
\newcommand{\wiii}{\mbox{$w_1$}}
\newcommand{\wii}{\mbox{$w_2$}}
\newcommand{\wi}{\mbox{$w_3$}}
\newcommand{\ca}{\ion{Ca}{2}}
\newcommand{\son}{\mbox{S/N}}
\newcommand{\vclus}{\mbox{$v$}}
\newcommand{\diag}{\mbox{$\alpha$}}
\newcommand{\kms}{\mbox{$\rm km{\ts}s^{-1}$}}
\newcommand{\ew}{\mbox{$EW$}}
\newcommand{\ews}{\mbox{$\Sigma Ca$}}
\newcommand{\pmu}{\mbox{$P_{\mu}$}}
\newcommand{\fe}{\mbox{[Fe/H]}}
\newcommand{\bv}{\mbox{$B-V$}}

\newcommand{\vv}{\mbox{$V$}}
\newcommand{\red}{\mbox{$E(B-V)$}}
\newcommand{\vhb}{\mbox{$V_{HB}$}}
\newcommand{\vclush}{\mbox{$\vclus_H$}}
\newcommand{\redew}{\mbox{$W^{\prime}$}}
\newcommand{\vhbmv}{\mbox{$V_{HB} - V$}}
\newcommand{\mAA}{\mbox{\rm \AA}}

\documentstyle[11pt,aaspp4]{article}

\lefthead{Rutledge et al.}
\righthead{Globular Cluster Ca II Metallicities}

\begin{document}

\title{Galactic Globular Cluster Metallicity Scale From the \ca\ Triplet \\
       I. Catalog}

\author{Glen A. Rutledge,  James E. Hesser\altaffilmark{1}, and Peter B. Stetson}
\affil{	National~Research~Council~of~Canada, 
	Herzberg~Institute~of~Astrophysics,
	Dominion~Astrophysical~Observatory, 
	5071 W. Saanich Rd., 
        RR5, Victoria, BC V8X~4M6, Canada 
	\\ Electronic mail: firstname.lastname@hia.nrc.ca}
 
\author{Mario Mateo\altaffilmark{1}}
\affil{Department of Astronomy, University of Michigan, 821 Dennison Bldg.
       Ann Arbor, MI 48109-1090 
       \\ Electronic mail: mateo@astro.lsa.umich.edu}
 
\author{Luc Simard}
\affil{Department of Physics and Astronomy, University of Victoria,
       P.O.~Box 3055, Victoria, BC V8W~3P6, Canada 
       \\ Electronic mail:  simard@beluga.phys.uvic.ca}

\author{Michael Bolte} \affil{UCO/Lick Observatory, University of
California,
	Santa Cruz, CA 95064 
	\\ Electronic mail: bolte@lick.ucsc.edu} 

\author{Eileen D. Friel} 
\affil{Maria Mitchell Observatory, 3 Vestal St. 
        Nantucket, MA 02554 
	\\ Electronic mail: efriel@mmo.org} \and
 
\author{Yannick Copin} 
\affil{Ecole Normale Sup\'erieure de Lyon, 46, allee d'Italie, 69364 Lyon
       cedex 07 
       \\ Electronic mail: Yannick.Copin@ens.ens-lyon.fr}

\altaffiltext{1}{Visiting Astronomer, Las Campanas Observatory.} 

\begin{abstract}

We have obtained 2640 CCD spectra with resolution $\sim$4~\AA\ in the
region 7250--9000~\AA\ for 976 stars lying near the red giant branches in
color-magnitude diagrams of 52 Galactic globular clusters.  Radial
velocities of $\sim$16 \kms\ accuracy per star determined from the spectra
are combined with other criteria to assess quantitative membership
probabilities.  Measurements of the equivalent widths of the infrared
calcium triplet lines yield a relative metal-abundance ranking with a
precision that compares favorably to other techniques.  Regressions
between our system and those of others are derived. Our reduction
procedures are discussed in detail, and the resultant catalog of derived
velocities and equivalent widths is presented.  The metal abundances
derived from these data will be the subject of a future paper.

\end{abstract}


\pagebreak[0]
\section{Introduction}

The absolute and relative ages of globular clusters in the Galaxy and in
the nearest Local Group galaxies provide unique constraints on cosmology
and early epochs of galaxy formation. However, the ages of globular
clusters cannot be determined, even in a differential sense, without
knowledge to high precision of their chemical composition (or
metallicity). An error of 0.3~dex in the overall heavy element abundance
of a cluster --- usually denoted by [Fe/H] --- corresponds to an error of
about 3~Gyr in the age derived from fitting an otherwise absolutely
correct isochrone to main-sequence photometry of perfect accuracy. Even
for some bright, nearby clusters, recent careful abundance measurements
differ by more than this amount, which reflects the challenges of detailed
analyses from stellar spectra.  Without reliable metallicity
determinations for many clusters, and especially for the crucial clusters
near the Galactic center, we cannot hope to test models of Milky Way
formation in a compelling fashion.

In 1989 we carried out a photometric and spectroscopic program at Las
Campanas Observatory that was aimed at developing a highly precise
relative ranking of globular cluster abundances from measurements of the
\ca\ triplet lines in the near infrared spectra of 12-15 probable red
giant members of each of 52 clusters. The early work of
\markcite{armzinn88} Armandroff and Zinn (1988, hereafter AZ88) used the
\ca\ triplet lines formed in the integrated light of Galactic globular
clusters, from which it appeared that an internal precision of 0.15~dex per star was
possible provided chromospherically active stars were avoided. Moreover,
by working in the infrared, sensitivity would be reduced to the high and
variable reddening towards many clusters of great interest for the
evaluation of formation scenarios for the Galaxy.  At the time the project
began, it was a relatively unexplored empirical approach which had been
applied primarily to integrated or composite light, and our goal was to
acquire data of such quality on individual giants that we could assess
thoroughly and independently the optimum procedures and relative merits of
this technique.

Since we undertook this project, several others have exploited with
great effect the \ca\ triplet technique applied to individual giants
for estimating abundances of globular clusters, with particular
emphasis upon distant and/or sparse objects (see, e.g., 
\markcite{armzinn88} Armandroff and Da Costa 1991 [hereafter AD91],
\markcite{olszewski91} Olszewski et  al. 1991, 
\markcite{arm92} Armandroff, Da Costa and Zinn 1992 [hereafter ADZ92],
\markcite{dacosta92} Da Costa, Armandroff and Norris 1992 [hereafter
DAN92], 
\markcite{suntzeff92} Suntzeff et al. 1992 [hereafter S92], 
\markcite{suntzeff93} 1993 [hereafter S93], 
\markcite{dacosta95} Da Costa and Armandroff 1995 [hereafter DA95],
\markcite{geisler95} Geisler et al. 1995 [hereafter G95], and 
\markcite{suntzeff96} Suntzeff and Kraft 1996 [hereafter SK96]). 
While these programs have provided
numerous results of widespread interest, the original motivation of our
program remains. In this paper we describe how we optimized our
reduction of the spectral data (\S \ref{excal}) to provide radial 
velocities (\S\ref{radvel}) and equivalent widths (\S \ref{EW}), compare 
our prescriptions and results with those of other workers
(\S \ref{transform}), and present a catalog of the individual stellar 
results (\S \ref{catalog}).  Following the AD91 prescription, the 
cluster reduced equivalent widths, \redew, are calculated 
(\S \ref{reduced_ew}).  A companion paper discusses the calibration
of our cluster \redew\ values to \fe\ values, and the astrophysical
implications of our results.

\pagebreak[0]
\section{Observations} \label{obs}

Spectra were obtained at the Las Campanas Observatory's 2.5m Dupont
telescope equipped with the modular spectrograph and the Canon 85mm f/1.2
camera.  A GG495 filter was used to block the second and higher spectral
orders.  The TI\#2 detector (800 $\times$ 800 thinned CCD; readout noise =
11~$e^{-}$~pix$^{-1}$; gain = 1.35~$e^{-}$ per ADU; scale =
0.85~$\arcsec$~pix$^{-1}$) was used with an 831~l~mm$^{-1}$
(8000~\AA\ blaze) grating, which produced a dispersion of
2.19~\AA~pix$^{-1}$ and spectral coverage from 7250--9000~\AA.  The
$8\arcmin \times 1.25\arcsec$ slit provided an instrumental spectral
resolution of $\sim$4~\AA.

Observations were obtained on two 1989 runs:  1) April 13--20 and 2) July
13--21.  Of the 52 clusters observed, 23 were observed during the first
run only, 26 were observed during the second run only, and three were
observed during both runs to check the consistency of our results.  In
each cluster, spectra were obtained for 10 to 20 stars selected from
published color-magnitude diagrams (CMDs) to lie on the red giant branch
(RGB) and, if proper motion data were available, to be likely
proper-motion  members.  Probable asymptotic branch (AGB) stars were avoided, as
were horizontal branch (HB) stars, and known variable stars near the RGB tip. 
Slit positions were chosen to contain at least two stars per spectrograph rotation.

Each star was observed two or three times consecutively, with an Fe-Ar arc
taken before and after each sequence for the wavelength calibration.
Occasionally the same star was observed on different nights, or with a
different slit orientation, to check for systematic effects in our
results. Exposure times for a single frame ranged from 2 min to 17 min.

The adopted data for the clusters we observed are presented in 
Table~\ref{clusters}, where the columns are, respectively:  1) the running
cluster identification number used in Figure~\ref{veldiff.grp}; 2,3,4) the
NGC, other cluster, and IAU names; 5,6) Galactic longitude and latitude in
degrees; 7) the visual magnitude of the horizontal branch level, \vhb;
8,9) the radial velocity and associated uncertainty, $\vclus_H$ and
$\sigma(\vclus_H)$; 10) the mean reddening for the cluster; 11) the central
velocity dispersion from \markcite{pryor93} Pryor and Meylan (1993,
hereafter PM93); 12,13) the metallicity and associated uncertainty of the
cluster taken from \markcite{zinn84} Zinn and West (1984, hereafter ZW84);
14) the standard deviation adopted for the \vv\ photometry, which is
used in \S \ref{reduced_ew} during the robust line fitting technique to
determine the reduced equivalent width, \redew, of the cluster (this value
is estimated from the scatter in the CMDs published by the authors from
which we adopted the photometry [see Appendix \ref{notes}], and
represents a combination of both the photometric errors and differential
reddening within the cluster).  The data from columns 5-10 were taken from
a 1994 version of the  \markcite{harris96} Harris (1996) electronic MWGC 
catalog (hereafter
referred to as the MWGC catalog), and references can be found in Appendix
\ref{notes}.

\placetable{clusters}

\pagebreak[0]
\section{Extractions and Calibrations} \label{excal}

Since there was not an overscan region on our detector, the bias level
of each frame was estimated from the mean level of the bias frames.
This was a satisfactory approach for run one, where the bias level
remained constant  at $\sim$550~ADU.  However, due to a CCD
electronics problem, the bias level in run two varied between 540 and
650~ADU on timescales of a few hours.

The illumination response along the slit resulted in a $\sim$21\%
reduction in transmission from one end of the slit to the other.  This 
effect was independent of slit rotation, and was found in both the object 
and flat field frames.  By normalizing the flat field frames along the 
dispersion axis only (with the IRAF\footnote{ IRAF is distributed by the 
National Optical Astronomy Observatories, which is operated by the Association 
of Universities for Research in Astronomy, Inc., under cooperative agreement
with the National Science Foundation.} task \bf response\rm), we could use
them to remove the illumination response.  For run two, where
the bias level of each frame was uncertain, the illumination response may
not have been removed correctly by this procedure. For these data we chose
sky windows on both sides of, and immediately adjacent to, the stellar
spectrum being extracted so that a low-order fit between the windows would
account satisfactorily for any residual errors in the illumination
response and bias level.  Spectra in crowded fields were not extracted
from run two data when windows appropriate for accurate sky subtraction
could not be identified adjacent to a spectrum.

We removed cosmic rays were removed with the IRAF task \bf cosmicrays, \rm 
while any remaining cosmic rays seen in a visual inspection of the two-dimensional
images we removed using IRAF's \bf imedit.  \rm We rectified residual distortion in
the images (manifested by curved night sky lines near the edges of the
frames) with the IRAF tasks \bf fitcoords \rm and \bf
transform. \rm

Several bad columns and pixels were noted before the observations were
taken, and stars were placed on the slit to avoid them.  Several other 
unreliable sections of the CCD were mapped and
avoided during spectral extractions.  Charge skimmed columns, in which
the percentage of electrons skimmed varied with time, were also
discovered during the reductions; no stars were extracted which fell on
these columns. 

Spectral extractions were made with IRAF's \bf apextract \rm tasks.  Two sky
windows, with a minimum of 15 pixels each, were chosen on either side of
the star, and a linear fit between the median values in the two windows
was used to define the sky level at the position of the stellar spectrum.
The windows were chosen to be as close to the star as possible, while
still adhering to the run two constraints mentioned earlier. To facilitate
the placement of the sky windows, a maximum of one bad column or charge
skimmed column was permitted to lie in a window and be dealt with by  the
medianing process.  The arc spectra were extracted with the identical
parameters used for the stellar extractions.  The two arc exposures
associated with a given star were averaged and the resultant digital
spectra were logically connected to the appropriate extracted stellar
spectrum.

We calculated the dispersion solution for each spectrum with a FORTRAN
program (similar to IRAF's World Coordinate System, which was unavailable
at the time of the reductions) that did not alter the pixel binning or the
pixel values, but rather wrote the coefficients of the
dispersion solution to the headers of the individual spectra.  For each
arc,  the program found every line above a threshold value and separated
from all other lines by at least two pixels, and fit each with a Moffat
function of exponent four to establish an accurate pixel center.  A
Legendre polynomial with five terms was fit to give the wavelength
dispersion solution.  This was done consecutively for all the spectra of a
given run, and the mean and $\sigma$ of the residual ($\lambda_{calc}  -
\lambda_{lab}$) for each line was calculated.  If the mean residual was
greater than 0.5~\AA, then the line was not used for the final dispersion
solution.  The remainder of the lines were weighted such that
\pagebreak[0] \begin{eqnarray} \nonumber w = 1.0 & if & \sigma \leq 
0.02~\mAA,  \\ \nonumber w = 0.02/\sigma & if & \sigma > 
0.02~\mAA.  \end{eqnarray} 
We applied these new weights in the final dispersion solution calculation
for each spectrum. In the end,  16 lines in  the
wavelength interval 7272~\AA\-- 8668~\AA\  were used.

To estimate the \son\ of each spectrum, we used two wavelength windows:
8580--8620~\AA\ and 8700--8800~\AA. These windows were chosen such that
none of the globular cluster \ca\ triplet lines would be velocity
shifted into them.  In each of these regions a robust line fitting
technique was used to fit a straight line to the pixel values, where
the absolute deviation was minimized in the fit rather than the square
of the deviation.  Let $N_i$ be the pixel value divided by its fitted
value, and let the mean and average deviation of all $N_i$ in a window 
be denoted as $N_{mean}$ and $N_{dev}$ (after clipping 
by 3 $\times$ $N_{dev}$).  The
\son\ for each window was then estimated to be $N_{mean}$/$N_{dev}$,
and the \son\ for the spectrum in the relevant \ca\ triplet region was
taken to be the average \son\ calculated for the two windows.  
Figure~\ref{soncomp.grp} shows a plot of the \ca\ triplet region for four of
our program spectra having \son\ values ranging from 12 to 125.  The
distribution of \son\ values for all of our program spectra can be
found in Figure~\ref{sondist.grp}.  The $\sim$1\% of the spectra with
\son\ $\lesssim$ 15 were not analyzed further.

\placefigure{soncomp.grp}
\placefigure{sondist.grp}

\pagebreak[0]
\section{Radial Velocities} \label{radvel}

Radial velocities aid in identifying {\it bona fide\/} cluster members,
particularly for those clusters projected against populous star fields.
While for membership assessment only relative radial velocities are
required, experimentation suggested that our data could be used for
independent velocity determinations. The procedures adopted are
described below.

\pagebreak[0]
\subsection{Cross Correlations} \label{cross}

We used a FORTRAN program to determine velocities by cross correlation 
against a template spectrum.  NGC 6809 star 2441 (II-4-41),
\son\ = 180, served as our template.  All spectra were continuum
normalized and rebinned to a log $\lambda$ scale.  The correlation
function, $C(\Delta\lambda),$ between the program spectra ($P$) and the
template spectrum ($T$) was calculated to be the sum of $T(\lambda)
\times P(\lambda - \Delta\lambda)$ between $\lambda_{template} =  8350$~\AA\ to
$8750$~\AA\  (which corresponds to $\lambda_{rest} \sim$8345~\AA\ to
8745~\AA) for a large range of $\Delta\lambda$ values.  These limits were chosen
to avoid telluric H$_2$0 features between $\sim$8100--8300~\AA, and
$\gtrsim$ 8800~\AA.  The maximum value of $C(\Delta\lambda)$ and the adjoining
$\pm$3 $\Delta\lambda$ values were then fit by a parabola whose center provided the
initial velocity estimate.  The program spectra were then Doppler
shifted by the initial velocity estimate and the cross correlation
repeated to get an additional velocity shift which ensured that the same
wavelength region in each spectrum was being used in the correlation;
this process was repeated until the velocity shift was stable to well
within our errors.  The final velocity was used to shift the spectra to
the template velocity, so that the band windows used in \S \ref{EW} to
calculate the \ew s were aligned properly.

\pagebreak[0]
\subsection{Velocity Errors} \label{vel_err}

Given a stable Cassegrain spectrograph insensitive to rotation angle
and changing gravity vector, the uncertainties in our velocities will
be dominated by slit centering errors. In order to achieve maximum
throughput as well as reliable relative velocities, considerable care
was spent in the slit rotation process to ensure that the prime pair of
stars was well centered on  the 1.25$\arcsec$ wide slit. This width
corresponded to 1.47 pixels on the image plane,  or 3.22~\AA\ 
$\sim$97~\kms\ at 8500~\AA. As shown below, our velocities per star appear to
be accurate to $\sim \pm 16$~\kms, after centering and other 
uncertainties are considered.

\pagebreak[0]
\subsubsection{Internal Errors} \label{vel_internal}

Our observational procedure ensured a large sample of stars for each
run that were observed at least twice consecutively with the same
exposure time. From these, the standard deviation of the velocity
measurement could be determined and compared to the mean \son\ of the
spectrum. The results are found in Figure~\ref{sonvel.grp}.  The median
standard deviation for run one, 7.7~\kms, and for run two, 8.0~\kms,  are
indicative of the internal precision and correspond to measuring shifts
between the program spectra and the template spectrum at the  $\sim$0.1
pixel level, which is typical for cross correlation techniques.

\placefigure{sonvel.grp}

\pagebreak[0]
\subsubsection{External Errors} \label{vel_external}

Several tests enable us to characterize the external accuracy of our
data; these include comparison of observations on different nights
within a run, observations on different runs, and observations with
different spectrograph rotations. We generally have three consecutive
spectra for every slit position, so comparisons below are made between
the median of each group taken under the different conditions.  The 
absolute value of the difference between two observations, each with
standard error $\sigma$, has an expectation value of $\sqrt{2} \sigma$;
this formulae was used below to estimate the standard error where appropriate.

Star 3204 in NGC 3201 (see Table~\ref{stars}) was observed
consecutively 19 times during run one at positions ranging over the
entire length of the slit. The standard deviation of the velocity was
14.2~\kms,  and no significant trend of derived velocity was found as a
function of position along the slit.  There were 223 and 37 stars
observed on more than one night of runs one and two, respectively;
there were no significant differences found from night to night, and
the standard errors derived from the mean absolute value of the star differences 
were $16 \pm 14$(s.d.)~\kms, and $12 \pm 12$(s.d.)~\kms,\ respectively.  
Fourteen and seven stars were
observed in runs one and two, respectively, with different spectrograph
rotation angles. The standard errors derived from the mean absolute differences 
were, respectively, 14$\pm$9~(s.d.)~\kms\ and 22$\pm$10~(s.d.)~\kms;  in neither 
case was a trend observed as a function of rotation angle.
From 15 stars observed in common between the two runs, the mean
difference between the velocities (run two$-$run one) was $-3.5 \pm$
10.1~(s.d.)~\kms.

An independent estimate of our uncertainties taking into account the
internal velocity dispersion of the clusters was made by comparing our
velocities with results from \S \ref{vel_results}.  Our calculated
dispersion, $\sigma_{calc}$, for each cluster for which we had 10 or
more stars was compared to the internal velocity dispersion,
$\sigma_{int}$, as given by \markcite{pryor93} Pryor and Meylan
(1993).  The mean excess in dispersion for 25 such clusters is
$\sigma_{e} = (\sigma^2_{calc} - \sigma^2_{int})^{0.5} =
16.0\pm$5.7~(s.d.)~\kms.

In summary, while the internal precision of an individual stellar
velocity appears to be $\sim$8~\kms, the more relevant external
uncertainties (arising from centering errors, flexure, etc.) are 
$\sim$16~\kms.  The velocity of each star relative to its cluster 
velocity given in the MWGC catalog is presented in Table~\ref{stars}.

\pagebreak[0]
\subsection{Template Velocity Zero Point and Cluster Velocities} \label{zero}

As noted earlier, star 2441 (II-4-41) of NGC~6809 (M55) was chosen as our
template for the cross correlations. Since we did not observe radial
velocity standard stars, its ex post facto choice was based upon it being
a relatively high \son\ observation of a globular cluster giant from a
cluster with a well determined radial velocity and velocity dispersion.
To produce velocities on the standard system we need to assign a velocity
to 2441. We could have chosen to use the MWGC catalog value (174.9$\pm$0.4
\kms) for the cluster, and ignore the possibility that this star might
have a detectable offset therefrom.  The latter possibility seems ruled
out by the unpublished measurements of Pryor and collaborators who used the
radial velocity scanner on the Canada-France-Hawaii Telescope, where the
velocity of 2441 was found to be 177.6$\pm$0.5 \kms\ relative to the
cluster mean velocity (for a 20 star dataset), 176.6$\pm$0.9 \kms, and a
cluster velocity dispersion of 3.8~\kms.  We chose, however, to set the
template velocity zero point by minimizing the difference between our
cluster velocity estimates and those given in the MWGC catalog, as we now
describe.

Our initial cluster velocity estimate was 
the median velocity after five iterations of 3$\sigma$ clipping.
Our final cluster velocity estimate also accounted for the central
internal velocity dispersion, $\sigma_{int}$, of each cluster. The
latter values, listed in Table~\ref{clusters}, are from
\markcite{pryor93} Pryor and Meylan (1993).  If the cluster was not
listed by Pryor and Meylan, then a typical value of 5~\kms\ was used.
Using our estimate of the external error in the measurement of the
velocity of a single observation of a single star, $\sigma_e =
16$~\kms\ (see \S \ref{vel_err}) for our measurement error, the
observed dispersion for each cluster should be $\sigma_{obs} =
(\sigma^2_{int} + \sigma^2_e)^{0.5}$.  The mean of all stars within
$3\sigma_{obs}$ of our initial velocity estimate form our final 
cluster velocity estimate, $\vclus$.   If these stars are drawn from a 
normal distribution with $\sigma = \sigma_{obs}$, then the variance in
$\vclus$ can be estimated as $\sigma^2(\vclus) \sim const^2 \times
\sigma^2_{obs}/N$, where $N$ is the number of stars entering the mean,
and $const$ is a constant that is determined below.

To obtain the template velocity zero point, we compared our velocity
determinations of 16 clusters for which the MWGC catalog quotes
velocity errors $<$1~\kms, and for which our estimate was based upon
$>$10 stars.  Let $\vclus_H$ and $\vclus$ be the catalog and our
values, respectively.  For each cluster, the difference, $\Delta \vclus
= \vclus - \vclus_H$, and the variance, $\sigma^2(\Delta \vclus) =
\sigma^2(\vclus) + \sigma(\vclus_H)^2$, were calculated (where the
value of $const$ in $\sigma^2(\vclus)$ was set to 1 in this analysis).
The velocity of the template, $v_{template}$,  was taken to be the
weighted mean ($w_i = 1/\sigma^2(\Delta \vclus_i)$) of $\Delta
\vclus_i$, 172~\kms.  The mean error of unit weight ($m.e.1$) was
calculated as follows:
  
\begin{displaymath}
m.e.1 = \left [\left( \sum \frac{(\Delta \vclus_i - v_{template})^2}
{\sigma^2(\Delta \vclus_i)} \right)/ {\nu}\right ]^{\frac{1}{2}} = 2.3,
\end{displaymath} where the sum is over the  16 clusters used and
$\nu=15$.  The uncertainty in our template velocity zero point is then
taken to be $\sigma(v_{template}) = m.e.1 \times (\sum 1 /
\sigma^2(\Delta \vclus_i))^{-0.5} = 2.2 \rm \; \kms.$ If we accept the
cluster velocity errors given in the MWGC catalog, then the $m.e.1$
value above indicates that  $const \sim2.3$, in the definition of
$\sigma(\vclus)$, will give a realistic external error estimate for
$\vclus$.  It is unclear why $m.e.1$ is 2.3 rather than 1; either the normal
error estimate, $\sigma_{obs}/\sqrt{N}$, is not appropriate for our
error distribution,  or the errors quoted in the MWGC catalog are 
underestimated (e.g., perhaps the quoted errors are more correct 
estimates of internal errors, rather than external errors).

In summary, we assign a velocity of $172 \pm 2.2~\kms$ to star 2441,
and our external uncertainties for the cluster velocities are,
$\sigma(\vclus) \sim2.3 \times \sigma_{obs}/\sqrt{N}$, 
where $N$ is the number of stars entering the mean.

\subsection{Velocity Results} \label{vel_results}

Our cluster results are presented in Table~\ref{cluster_res}, where
the columns are respectively,
1,2,3) as in Table~\ref{clusters};
4,5,6) our mean cluster velocity, $\vclus$, external error, $\sigma(\vclus)$, 
and the number  of stars used to estimate the mean velocity, $N$; 
7,8) the difference $\Delta \vclus = \vclus - \vclus_H$, and 
$\sigma(\vclus - \vclus_H) =({\sigma}^2(\vclus) + \sigma^2(\vclus_H))^{0.5}$,
where the $\vclus_H$ values can be found in Table~\ref{clusters};
9,10,11) the reduced equivalent width, \redew\ of the cluster, its 
associated uncertainty, $\sigma(\redew)$, and the mean error of unit 
weight, m.e.1, in the fit of the cluster as described in \S \ref{reduced_ew}. 
The difference between each star's velocity and the cluster velocity given in 
the MWGC catalog is plotted in Figure~\ref{veldiff.grp}, and the individual
stellar velocities are tabulated in Table~\ref{stars} in \S \ref{catalog},
below.

\placefigure{veldiff.grp}
\placetable{cluster_res}

While the goal of this project was not to determine accurate cluster
velocities, it is reassuring that our cluster velocity
estimates generally agree well with the estimates given in the MWGC
catalog, with only five clusters having differences greater than
3$\sigma$: NGCs 2298, 5897, 6101, 6553, and 6981.  For NGCs 2298, 5897
and 6101, \markcite{geisler95} G95 recently reported
velocities of $150.4\pm1.3$~\kms, $102.9\pm1.0$~\kms and
$364.3\pm1.9$~\kms, respectively.  In all three cases the differences
between our velocities and theirs are less than $1\sigma$.

Finally, NGCs~6235, 6528 and 6681 presented particular challenges when
identifying stars to define the initial velocity estimate.  For
NGC~6235, there was no obvious grouping of velocity measures in our
sample; therefore, the catalog velocity was used as the first
estimate of this cluster's velocity and only five stars satisfied the
iterative clipping procedures to determine our velocity, which
differs from the catalog value by $0.1\sigma$.  NGC~6528, with (\em 
l,b\rm) = ($1.1 \arcdeg$,$-4.2 \arcdeg$), is projected onto the dense
star fields of the Galactic center.  However its high radial velocity
made it reasonable to reject all the stars with velocities near 0~\kms,
which are most likely bulge stars, when determining the initial
estimate; eight stars survived the clipping to enter our final
velocity, which differs from the catalog value by 3.4$\sigma$.
However, membership remains something of a concern for any cluster with
such contamination problems.  Similarly, NGC~6681 is close to the
bulge, but has a high radial velocity.  Since there was a group of
stars close to the velocity quoted in the catalog, and a scattering
of stars with velocities closer to 0~\kms, the median velocity of the
five stars close to the catalog velocity was used as a first estimate
of this cluster's velocity, which produced a final cluster velocity
estimate that differs from the catalog value by
0.1$\sigma$.  Due to these problems, the cluster velocities we obtain,
and our assignment of membership probabilities for stars from these
three clusters, are less reliable than those for the other 49
clusters.

\pagebreak[0]
\subsection{Cluster Membership Probabilities} \label{vel_mem}

For each cluster, a Gaussian probability distribution function (PDF)
was defined, $P_G(v_{star},\vclus,\sigma)$, which represents the
probability that a star with velocity $v_{star}$ was drawn from the
assumed cluster Gaussian velocity distribution with mean $\vclus$ and
$\sigma = [\sigma^2_{obs} + \sigma^2(\vclus)]^{0.5}$ as defined in 
\S \ref{zero}. A field star population was defined by selecting all
stars which differed from their cluster velocity by more than
$2\sigma$.  Of the 976 stars in our sample, 158 fit this
criterion.
These presumed field stars form a symmetrical distribution about $v =$
0~\kms, and are satisfactorily fit by a Gaussian with $\sigma_{field}$
= 75~\kms.  Therefore, the field PDF was defined as
$P_G(v_{star},0,\sigma_{field})$.  To define accurately the probability
of a star being a velocity member of a given cluster, it is also
necessary to know the relative number density of cluster stars and
field stars for each cluster at the radii where we observed.  To first
order, this number can be estimated as the total number of cluster
stars in our sample over the total number of field stars, as defined
above (i.e. $N_{c/f} = 818/158$). The probability that a star is a
cluster member on the basis of its velocity relative to the cluster
mean is, \begin{displaymath} P_v = \frac{ N_{c/f}
P_G(v_{star},\vclus,\sigma_{obs})} {N_{c/f}
P_G(v_{star},\vclus,\sigma_{obs}) + P_G(v_{star},0,\sigma_{field})}.
\end{displaymath} This approach to assigning individual star membership
probabilities offers at least two advantages compared to simple
$\sigma$-clipping.  First, it allows for the greater ease of
distinguishing between field and cluster stars when the cluster has a
high velocity.  Second, a star whose velocity deviates from the cluster
mean in the direction of $v = 0$~\kms\ is more likely to be a field
star than one that deviates in the opposite direction.  These two
effects are  reflected in our probability scheme since the field star
population is centered on $v = 0$~\kms, and the cluster population is
centered on $\vclus$.  Membership probabilities thus calculated are
listed for each star in Table~\ref{stars}.

\pagebreak[0]
\section{\ca\ Triplet Equivalent Widths} \label{EW}

Measurement of  the \ew\ of an absorption feature is usually done by
defining continuum bandpasses on each side of the feature, and linearly
interpolating the average or median intensities in each of the
bandpasses to define the continuum at the feature wavelength.  The
\ew\ is then the integral over the feature bandpass of the difference
between the continuum and the feature (see \S \ref{bandpass}).  The
feature is defined either directly by the spectral intensities or by
some analytical function fitted to them (see \S \ref{line_fit}).  For
\ca\ triplet work, it is also necessary to combine the three lines in
some manner to get a net \ca\ index, which we will denote as \ews\ 
(see \S \ref{ew*}). Table~\ref{ew_tech} summarizes the approaches used by 
previous practicioners of \ca\ triplet work, where the columns are,
respectively:
1) the referenced paper;
2) the method used to combine the three \ca\ triplet lines into a
single \ca\ index for the star, \ews;
3) the method used to define the line feature in the spectrum;
4) the reference for bandpass limits used to define the continuum and 
feature regions.  Table~\ref{bands} defines the
bandpasses that were adopted by other authors  and by ourselves\footnote{ Note
that the instrumental resolution (usually defined by the FWHM of the 
arc lines) was $\sim$3~\AA\ for the earlier
studies, with values ranging from
2.5~\AA\ (\markcite{olszewski91}Olszewski et al. 1991) to 4.8~\AA\
(\markcite{armzinn88}Armandroff and Zinn 1988), compared to our
3.4--4.8~\AA\ (see \S~\ref{bandpass}).}, where the columns are, respectively:
1) the paper where the bandpasses are defined;
2) the \ca\ triplet line name defined by its rest wavelength;
3) the line center adopted by other authors and derived by ourselves;
4) the limits of integration for the line;
5,6) the limits used to define the continuum on the blue and red 
side of the line.

\placetable{ew_tech}
\placetable{bands}

Since there were many multiply observed stars in our sample, a series
of tests were performed to determine the optimum method of calculating
\ews.  In these tests, the mean, $<\ew >,$ and standard deviation,
$\sigma(\ew ),$  of the \ew\ was calculated for each line in every star
that was observed at least twice consecutively with the same exposure
time\footnote{In \S \ref{line_fit} and \S \ref{bandpass}, the sample
was also restricted to stars with successful \ew\ calculations for
all three line fitting methods to avoid bias arising from one method
having a higher success rate than another.  For the remaining tests, we
restricted the samples to stars which had successful \ew\ calculations
from the Moffat line fitting technique for all three \ca\ triplet
lines, which was necessary for our \ews\ to be calculated.}.  A
diagnostic, \diag,\ defined as the median $\sigma(\ew )$ divided by the
median $<\ew >$, provides essentially an estimate of the inverse
\son\ for the \ew .  A technique was then developed to minimize \diag.
Table~\ref{ew_diag} presents the \diag\ values for the various tests
described below, where the columns are respectively,
1) the technique used to calculate the equivalent widths;
2) the run that the test was performed on;
3,4,5) the \diag\ values for each of the three \ca\ triplet lines.

\placetable{ew_diag}

\pagebreak[0]
\subsection{Line Fitting Technique} \label{line_fit}

Three different techniques were tested for measuring equivalent widths:  
direct numerical integration, fitting the line with a Gaussian function, and 
fitting the line with a Moffat function of exponent 2.5.  Initially the 
\markcite{armzinn88} AZ88 line bandpasses were used, and for the continuum  
bandpasses the \markcite{armzinn88} AZ88 definition was used for \caiii, and the 
\markcite{arm91} AD91 definition was used for \caii\ and 
\cai\ (see Table~\ref{bands}), which we will refer to  as the AZ88/AD91
continuum bandpasses\footnote{
AD91 significantly modified the continuum bandpasses defined by AZ88 for \caii\ and
\cai\ to optimize their equivalent width calculations.
}.  
Between both runs, there was
a total of 600, 764, and 750 stars used for \caiii, \caii, and
\cai,\ respectively.  The first six rows of Table~\ref{ew_diag} show
that compared to the Numerical method, the Gaussian method shows an 
improvement in \diag\ of $\sim$10\% on average, and compared to the 
Gaussian method, the Moffat method shows an improvement of 
$\sim$4\% on average.  In addition, when plotting the deviations from 
the fitted profile for \em all \rm our spectra, it is clear that
the Moffat fit shows no obvious systematic differences from the observed
profile \footnote{
The exponent for the Moffat function was chosen to be 2.5 since, from
a visual inspection, it was found to fit our observed line profile the best.
}
, except for slightly underestimating the wings.  On the other hand, the
Gaussian fit underestimates the depth of the line by $\sim$5\%, overestimates
the FWHM by $\sim$10\%, and grossly underestimates the depth of the wings.
We have therefore adopted the Moffat fitting technique for the remainder 
of our analysis.  To quantify the differences between the line fitting 
techniques, we performed a spectrum-by-spectrum comparison for 1866 spectra
to relate the Ca index (see \S~\ref{ew*}) calculated  
with the Numerical method, \ews (N), the Gaussian method, \ews (G), and
the Moffat method, \ews (M).  A least squares fit results in,

\begin{displaymath}
\ews (N) = 1.005 (\pm 0.003) \cdot \ews (M) - 0.047 (\pm 0.013)~~rms = 0.12~\mAA
\end{displaymath}
\begin{displaymath}
\ews (G) = 0.995 (\pm 0.003) \cdot \ews (M) - 0.084 (\pm 0.005)~~rms = 0.04~\mAA,
\end{displaymath}
which indicates that there are slight zero point shifts between the techniques,
but the slopes are consistent with unity.  The larger $rms$ value for the
Numerical technique is consistent with its larger \diag\ value (\em c.f.~\rm
Table~\ref{ew_diag}).

\pagebreak[0]
\subsection{Line and Continuum Bandpass Windows} \label{bandpass}

Next,  we optimized the continuum bandpasses for the line
bandpasses of Armandroff.  Our continuum
bandpasses, listed in Table~\ref{bands}, were chosen to be as
large as possible, while still avoiding the telluric H$_2$O features
between $\sim$8100--8300~\AA\ and $\gtrsim$ 8800~\AA.  As seen in rows 7
and 8 of Table~\ref{ew_diag}, our new continuum windows reduced
\diag\ by $\sim$9\% for the \caiii\ line, while the \caii\ and 
\cai\ lines were insignificantly affected.  We  quantified the 
differences between the two
sets of continuum windows by making a star-by-star comparison for 764 
stars. This allowed us to relate the Ca index calculated with the Armandroff 
continuum windows, \ews (A), and with our larger continuum windows, \ews (L).
A least squares fit results in,
\begin{displaymath}
\ews (A) = 1.071 (\pm 0.004) \cdot \ews (L) - 0.19 (\pm 0.02)~rms= 0.13~\mAA,
\end{displaymath}  
which indicates that a small scaling factor and zero-point offset 
exists between the two methods.  All stars that lie significantly
off this relation were previously noted to have strong TiO absorption,
and were not further included in our analysis.

Finally, we adopted the new continuum bandpasses and tested a variety 
of line bandpasses that ranged from $\pm~3.5$~\AA\ to
$\pm~10$~\AA\ about each line center.  Our line fitting technique
simultaneously fit the amplitude, function-width parameter, and central
wavelength, $\lambda_c$, for each line of each spectrum.  The median
$\lambda_c$ for each line from all spectra are listed in Table~\ref{bands}, 
which are almost identical to the laboratory wavelengths defined
for these lines.  The feature bandpasses were defined around these
median $\lambda_c$ values.  In general, we found, as one might expect,
that \diag\ decreases monotonically from the larger to the smaller
bandpasses. The effect is strongest for \caiii, which ranges from 0.082
($\pm~10$~\AA) to 0.063 ($\pm~3.5$~\AA) for run one, and 0.091
($\pm~10$~\AA) to  0.074 ($\pm~3.5$~\AA) for run two.  For bandpasses
$\lesssim\pm4.5$~\AA, the change in \diag\ becomes insignificant. Thus,
for each line, the largest bandpass for which there was no
significant increase in \diag\ from the $\pm~3.5$~\AA\ band was analyzed
further.  The
values chosen for \caiii, \caii~and \cai, $\pm~4$~\AA, $\pm~4.5$~\AA,
and $\pm~4.5$~\AA, respectively,  are about half the size of the
original Armandroff line bandpasses.  The \diag\ values for these line
bandpasses represent a 16.5\% reduction, on average, from the Armandroff 
line bandpasses. 

Although smaller line bandpasses significantly decrease the \diag\ values, 
they also dramatically increase sensitivity to changes in the 
instrumental profile, which in turn, induce systematic errors in our 
calculated equivalent widths.  Unfortunately, from an analysis of the 
arc lines associated 
with each spectrum, the FWHM were essentially constant for run one at 
$\sim$3.4~\AA, while for run two, the FWHM was $\sim$4.9~\AA\ for night one, 
$\sim$4.4~\AA\ for nights two to six, and $\sim$3.8~\AA\ for the remainder 
of the nights.  The changes in run two are attributed, \em ex post facto\rm, 
to changes in the spectrograph focus.
To quantify the effects of these changes in the FWHM of the intrumental profile
on our choice of line and continuum bandpasses,
a group of 532 pairs of spectra were collected, such that
i) the spectra were of the same star, and ii) the FWHM of the arc lines
associated with the two spectra differed by more than 0.1~\AA\ (the maximum
difference was 0.8~\AA).  Let $\Delta EW$ be the difference between \ews\ calculated
from the spectra with the larger arc FWHM and the smaller arc FWHM, and let 
$|\Delta FWHM|$ be the absolute value of the difference between the FWHM of each
of the arc lines.  When we use the AZ88 line bandpasses and our continuum bandpasses
(see the TP bandpasses in Table~\ref{bands}), a least squares fit results in,
\begin{displaymath}
\Delta EW = 0.008 (\pm 0.10) \cdot |\Delta FWHM| - 0.04 (\pm 0.03),
\end{displaymath}
so no significant trend is present.  However, when we make the same comparison 
with the the narrow line bandpasses and our continuum windows, we obtain, 
\begin{displaymath}
\Delta EW = -0.57 (\pm 0.07) \cdot |\Delta FWHM| +0.05 (\pm 0.02),
\end{displaymath}
which clearly shows that the \ew\ of a line is underestimated as
the FWHM of the instrumental profile increases, as one would expect.
Had our instrumental resolution remained constant throughout the two runs,
it is clear from the final \diag\ diagnostic in Table~\ref{ew_diag} that 
the narrower line bandpasses derived herein would have represented 
improvements over the AZ88 values.  With the data available, however, it
is evident that the larger line bandpasses are preferable,
and thus they were adopted for the remainder of our work.  Our final 
bandpasses (TP) are presented in Table~\ref{bands}.

\pagebreak[0] 
\subsection{Combining the \ca\ Triplet Lines: \ews} \label{ew*}

In their original work, \markcite{armzinn88} AZ88 measured 
\ca\ indices from integrated cluster light and they defined \ews\ to be
the sum of the three triplet lines.  When \markcite{arm91} AD91 used 
individual cluster giants, they determined that inclusion of the \caiii\ 
line added more noise than signal to \ews, and thus excluded it from the 
sum.  Since we have a large sample of multiply observed stars, we revisited 
this issue.

Our goal is to determine a weighted mean for \ews, \ews\ = \wiii\ewiii
+ \wii\ewii + \wi\ewi, where the weights minimize \diag\ for \ews.
For each line in each star in our sample we calculated the mean,
variance and covariance, ($m1, m2, m3$), ($v1, v2, v3$), and ($c12,
c13, c23$), respectively.  The median of these values was calculated
for a sample and respectively denoted as $m1_0, m2_0, m3_0, v1_0, v2_0,
v3_0, c12_0, c13_0,$ and $c23_0$.  A value equivalent to \diag\ defined
earlier was then calculated as 
\begin{displaymath} \tilde{\diag} =
\frac{(\wiii ^2 v1_0 + \wii ^2 v2_0
+ \wi ^2 v3_0 + 2 \wiii \wii c12_0 + 2 \wiii \wi c13_0 + 2 \wii \wi
c23_0)^{\frac{1}{2}}}{\wiii m1_0 + \wii m2_0 + \wi m3_0}. 
\end{displaymath} Even though the covariances
were less than the variances by at least a factor of ten, they were
retained. We proceeded as follows. We set the value of \wii\ to 1, and
we minimize $\tilde{\diag}$ with a downhill simplex method 
(\markcite{press92}Press et al. 1992)
by letting \wiii~and \wi\ vary.  For observing run one, the minima found
were \wiii\ = 0.55 and \wi\ = 0.62, while for run two they were
\wiii\ = 0.48, and \wi\ = 0.68.  The average, rounded to the nearest
tenth, for each weight was taken to give final weights of \wiii\ = 0.5,
\wii\ = 1.0, and \wi\ = 0.6.  These weights yield a final
\diag\ value for \ews~of 0.032 for run one, and 0.034 for run two.
If, on the other hand, both \wi\ and \wii\ were 1.0 and \wiii\ was 0.0,
the respective \diag\ values would be 0.035 for run one, and 0.037 for run
two. It thus appears that in our data, inclusion of the \caiii\ feature 
increases the \son\ of \ews, and we therefore include it with the
weights just defined.

\pagebreak[0]
\subsection{\ew\ errors} \label{ew_err}

In order to combine observations with varying \son\ in the most effective
way, and to measure the uncertainty in \ews\ for single observations, the
standard deviation in the \ews, $\sigma$(\ews), was computed for 594 stars that
were consecutively observed with identical configurations and exposure
times, and compared to the mean signal to noise, $<$\son$>$, for each star.  These 
values are plotted in Figure~\ref{sonmofsig.grp}.  The solid line overplotted on 
these data represents our estimate of the single observation internal $\sigma$
as a function of the \son\ of the spectra, and was calculated as follows.
A running median over 20 stars for both $\sigma$(\ews) and $<$\son$>$ was calculated,
and a quadratic fit was performed which is taken to be the error estimate for $<$\son$>$ less
than 100, and for $<$\son$>$ greater than 100, a constant 
$\sigma(\ews) =  0.067~\mAA$ was assumed. We assume our errors are approximately Gaussian, and weight 
the observations as $1/\sigma^2$ when calculating the mean \ews\ for a star that was 
observed multiple times.  The $\sigma$ listed for each star in Table~\ref{stars} is the 
error in the weighted mean, $[\sum 1 /\sigma^2(\ews_i)]^{-0.5}$, where the sum is over 
the number of spectra per star.

\placefigure{sonmofsig.grp}

A number of tests for systematic effects in \ews\ determinations were
possible, which also allowed us to estimate the external errors in
\ews. From 15 stars observed in common between runs one and two, the
mean difference in \ews~was found to be \ews(run two)$-$\ews(run
one)$=0.007\pm0.15$(s.d.)~\AA.  NGC 3201 star 3204 was observed consecutively 
19 times at different slit positions.  The standard deviation in \ews\ was 
0.17~\AA\ for a mean \son\ of 62, which  is $\sim1.4 \times
\sigma(\ews)$; no trend in \ews\ was seen as a function of
position on the slit.  There were 223 and 37 stars observed on
different nights of run one and run two, respectively.  No significant
offsets were found between any of the nights, and the standard errors derived
from the overall absolute value of the differences between nights was $0.12 \pm 
0.12$~\AA\ $[\son\ = 61 \pm 23; \sim1.0 \times \sigma(\ews)]$ and $0.09
\pm 0.07$~\AA\ $(\son\ = 71 \pm 15; \sim0.9 \times \sigma(\ews))$ for runs one and two, 
respectively.  Fourteen
stars from run one and seven stars from run two were observed with
different spectrograph rotation angles. The standard errors derived from the 
mean absolute value of the difference in \ews\ were $0.16 \pm$0.09(s.d.)~\AA\ and $0.17
\pm$0.10(s.d)~\AA, respectively.  For their respective mean
\son\ of 53 and 43, these correspond to error excesses of $\sim1.1
\times \sigma(\ews)$ and $\sim1.0 \times \sigma(\ews)$.

In summary, our external EW errors appear to be  consistent with our internal errors. 

\pagebreak[0]
\subsection{Transformations} \label{transform}

We have derived transformations between our \ews\ system, TP,  and those of
others. Observations of individual stars in common with various sources
are plotted in Figure~\ref{ewcomp.grp}.  The error bars for the abscissa
represent our external error estimates, while the error bars for the
ordinate represent the errors quoted by the other authors.  The dotted 
lines on
the graph represent a one to one correlation, while the solid line represents
our least squares fit to the data, which allows for
errors in both directions (\markcite{stetson89ls}Stetson 1989).  The 
regression coefficients for the fit $\ews(other) = m \cdot \ews(TP) + b$, 
are listed in Table~\ref{regression}, where
the columns are respectively, 1) the data source; 2,3) the slope of the
fit and its corresponding uncertainty; 4,5) the intercept of the fit and
its corresponding uncertainty; 6) the mean error of unit weight for the
fit; 7,8) the approximate minimum and maximum \ews(TP) for which the
regression is valid; and 9) the number of stars in common, N.  The $m.e.1$ value was 
calculated as, $m.e.1 =\left( \sum
\epsilon^2/\sigma^2 \right)^{0.5}\ts \nu^{-0.5}$, where the sum is over
all the stars used, $\epsilon$ is the deviation along the ordinate of a
star from the fitted relation, $\sigma$ is calculated for each star as
$(m^2\ts \sigma^2[\ews(TP)] + \sigma^2[\ews(other)])^{0.5}$, and $\nu$ is
the number of stars used minus two.  The uncertainties listed for the
coefficients are the formal uncertainties from the fit multiplied by the
$m.e.1$ value for the fit.  The fit to the SK96 data was done excluding 
the star with the largest \ews, NGC 104 L5622, since this star has weak
TiO bands present, and SK96 suggest that the line strengths may be varying 
in this star. 

\placefigure{ewcomp.grp}
\placetable{regression}

If we assume that our estimates for $\sigma(\ews)$ reflect our true external 
errors, then the $m.e.1$ values for S92, S93, G95, and SK96, which are 
greater than one, suggest that they have underestimated their errors.
For G95, the $m.e.1$ value reduces from 1.75 to 1.18 if the two most
deviant points are removed. 
DA95 compared their \ews\ values to those of other authors as well, but 
did not publish regressions.  Their results are qualitatively 
in agreement with ours, which suggests that the DA95 and S93 
\ews\ values are on the same system, the AD91 \ews\ values show a
positive slope with respect to the DA95/S93 system (i.e., the
difference between AD91 \ews\ and DA95/S93 \ews\ becomes larger
as \ews\ increases), and the ADZ92 \ews\  show a small 
excess of $\sim0.25~\mAA$ over the DA95/S93 \ews\
for the small range of \ews\ in common.  DA95 found
a small positive slope between the DAN92 \ews\  and the 
DA95/S93 system, whereas we find that all three systems are 
essentially the same. 

We attribute the differences in slope and zero point between studies to a 
combination of continuum window definitions (see \S \ref{bandpass}), 
line fitting techniques (see \S \ref{line_fit}), \ews\ definitions
(see \S \ref{ew*}), and slight changes in the intrumental resolution
and throughput properties of the spectrograph and detector used
(see \S \ref{bandpass}).

\pagebreak[0]
\section{Cluster Reduced Equivalent Widths: \redew} \label{reduced_ew}

As demonstrated by AD91, the change in \ews\ as a function of \vhbmv\ 
for stars within a cluster has a constant slope, $\Delta (\ews)/\Delta 
(\vhbmv)$, for all clusters. 
This result was corroborated  by subsequent studies for stars with
$\vhbmv \gtrsim 0$, where the mean slopes from the various papers are
presented in Table~\ref{m_author};  the columns are respectively,
1) paper reference,
2) mean slope from the paper, and
3) the uncertainty in the slope.
In most cases, a standard slope of 0.62~$\mAA\ts{\rm mag}^{-1}$ was adopted, 
which allows for the calculation of a reduced equivalent width, \redew,  
for a cluster, where the \ews\ of each star in the cluster is corrected 
to the level of the horizontal branch by subtracting 0.62(\vhbmv), and the 
mean of all cluster members is taken.  Since our method of 
calculating \ews\ 
differs from the previous studies,
we do not expect, \em a priori\rm, that our slope 
will be the same as theirs.  Since SK96, and S92 showed that the slope reduces
to $\sim$0.35 for RGB stars with $\vhbmv \lesssim 0$, we have calculated \redew\
for each cluster excluding these stars, which represents only $\sim$4\% of our total
sample.  We have also excluded stars which appear to be AGB stars, HB stars, have an
uncertain ID, or lie significantly off the locus defined by the other stars in the 
\ews, \vhbmv\ plot.  Furthermore, the weight of a star in the fitting procedure
is dependent on its radial velocity and proper motion, as described below. 

\placetable{m_author}

The technique we chose to calculate \redew\ for each cluster uses
an iterative,  robust algorithm to fit one slope and 52
intercepts simultaneously to all our data.  The intercepts represent
the value of \ews\ at the level of the horizontal branch for each
cluster, and thus are equivalent to \redew.  The analytic function fit
to our data by an iterative least squares technique, allowing for
errors in both directions, is $y_{i,j} = mx_{i,j} + b_j$, where 
$y_{i,j}$ is the \ews\ value of star $i$ in cluster $j$; 
$x_{i,j}$ is the \vhbmv\ value of star $i$ in cluster $j$; m is
the slope, $\Delta (\ews)/\Delta (\vhbmv)$, which is assumed to be
constant for all clusters; and $b_j$ is the intercept or reduced
equivalent width, \redew, of cluster $j$.
The errors, $\sigma_x$,  assigned to the $x_i$ values were assumed to
be constant for a cluster, and are listed in Table~\ref{clusters} as
$\sigma(\vv)$. They represent our estimate of the errors in \vv\ from a
combination of photometric errors, and errors due to differential
reddening within the cluster.  The errors, $\sigma_y$, assigned to the
$y_i$ values are our estimates of the external errors in the \ews\ of
the star, and are listed in Table~\ref{stars} as $\sigma(\ews)$.

The weight of each star 
in the fit is $w = f_p\ts f_r\ts \sigma^{-2}$, where the final 
$w, f_p,$  and $f_r$ are 
listed for each star in Table~\ref{stars}.  The $f_p$ value for a star 
is a constant throughout the iterative fitting, and is calculated as,
\begin{eqnarray}
\nonumber
f_p = P_{v}\phm{0} & if & P_{\mu} \rm~is~not~available,  \\
\nonumber
f_p = P_{v}\phm{0} & if & P_{\mu} > 0.2,  \\
\nonumber
f_p = 0\phm{P_{v}} & if & P_{\mu} \leq 0.2,  
\end{eqnarray}
where the $P_{\mu}$ and $P_v$ values for each star are the probabilities
of membership based on proper motions and velocities (see \S 
\ref{vel_mem}), respectively 
(listed in Table~\ref{stars}).  Since stars were selected to be close to the RGB of
the cluster, and are found as close to the cluster center as the published
photometry, finder charts, and stellar density permitted,
(both of which increase their probability of being cluster members), 
the proper motion probabilities were used conservatively to assess 
membership.
The $\sigma$ attached to each star is calculated as 
$\sigma = (m^2_0\ts \sigma^2_x + \sigma^2_y)^{0.5}$, where $m_0$ is
the estimate of the slope from the previous iteration.
The $f_r$ value assigns lower weight  to stars which lie
significantly off the locus defined by the other stars in the 
cluster, and is calculated as,
\begin{displaymath}
f_r =   \frac{1}{1 + [|\epsilon|/(\gamma\ts \sigma\ts m.e.1)]^{\beta}},
\end{displaymath}
where $\epsilon$ is the star's \ews\ deviation from its cluster
fit in the previous iteration; $\sigma$ is calculated as described
above; $\gamma$ and $\beta$ are constants discussed below; and 
$m.e.1$ is the mean error of unit weight for the cluster which is 
calculated as 
$m.e.1 =\left( \sum \epsilon^2/\sigma^2 \right)^{0.5}\ts \nu^{-0.5}$,
where the sum is over the number of stars in the cluster with 
$f_p \geq 0.75$, and $\nu$ is one less than this.  The constants 
$\gamma$ and $\beta$ were set to 3 and 4, respectively, which assigns 
equal values of $f_r$ to all stars within $\sim2\ts \sigma\ts m.e.1\ts$ 
of the cluster fit,  while $f_r$ drops rapidly to 1/2 when the star is 
$\sim3\ts \sigma\ts m.e.1\ts$ from the cluster fit.  The iterations 
were continued until there was no significant change in $m.e.1$ for any 
cluster.  For each cluster, the final intercept, \redew, its 
uncertainty, $\sigma(\redew)$, and the  $m.e.1$ value are listed in 
Table~\ref{cluster_res}, where $\sigma(\redew)$ is the formal uncertainty 
in the intercept from the fitting technique multiplied by $m.e.1$ for the cluster.
An $m.e.1$ value greater than one indicates that the scatter about the best fit
line is larger than what would be expected from the adopted errors in \ews\ and 
$V$ alone.  Since we are generally confident in our error estimates for \ews\
(see \S \ref{ew_err}), the
excess scatter in most clusters is likely due to underestimating the 
errors in $V$, including non-cluster members in our sample, including 
non-RGB stars in our sample, differential reddening within a cluster,
or, as a remote possibility, a Ca 
spread in the cluster.  The inhomogeneous nature of the photometry used to 
establish the ordinate in Figure~\ref{ewcmd.grp} seems likely to be the 
dominant source of scatter.

The slope that we obtain from this technique is 
$0.64 \pm 0.02~\mAA\ts{\rm mag}^{-1}$, with $m.e.1 = 1.6$, where the 
uncertainty quoted is the formal uncertainty from the fitting technique
multiplied by $m.e.1$.  The $m.e.1$ value here is calculated as it was for
individual clusters, except the sum is over the number of stars in 
\em all \rm  clusters with $f_p \geq 0.75$, and $\nu$ is this number minus 
53 (i.e., the number of clusters used plus one).  This slope is consistent
with the slopes that previous authors have found, even though our 
transformation results (\S \ref{transform}) suggest that our slope should
be slightly shallower.  This is most likely due to the fact that other
authors have not allowed for errors in $V$ when calculating the slope,
which causes the fitting technique to underestimate the true slope 
(\markcite{stetson89ls}Stetson 1989).  This method of calculating the mean
slope is more effective than simply taking the mean of slopes fit independently,
since clusters with stars having a small range in \vhbmv, and
therefore possessing little slope information, simply add noise to the latter
method, whereas they do not affect the former. The final fit for
each cluster is plotted in Figure~\ref{ewcmd.grp}.

We performed two experiments to investigate whether the slope, 
$\Delta (\ews)/\Delta (\vhbmv)$, is a function of \fe.
In the first, every cluster was fit by our technique independently,
and the resulant slopes of each cluster were plotted against their
respective \fe\ from ZW84, which are listed in Table~\ref{clusters}.  
The mean slope was $0.62 \pm 0.02~\mAA\ts{\rm mag}^{-1}$, and no 
significant trend with \fe\ was observed, although there was some indication
that the slope may become steeper for the more metal-rich clusters.  
To examine further any possible \fe\ dependence of the slope,
the clusters were split into four metallicity
bins, with each subset analyzed following the precepts described above
for the full sample. The results are presented in Table~\ref{m_zbin},
where the columns are respectively, 1) the \fe\ range as defined by ZW84
(see Table~\ref{clusters}), 2) the number of clusters in that \fe\ range,
3,4) the slope, $\Delta (\ews)/\Delta (\vhbmv)$, and uncertainty in the 
slope, derived for the clusters in the bin, and 5) the $m.e.1$ value
for the fit.
As seen in Table~\ref{m_zbin}, none of the slopes for the metallicity bins
differs significantly from the $0.64 \pm 0.04~\mAA\ts{\rm mag}^{-1}$ found
for the whole sample, nor is there compelling evidence that more metal-rich
clusters differ in this regard.  Accordingly, we will continue to assume that
the slope is constant for all metallicities.  With better photometry and 
membership information, this point would be well worth revisiting in the 
future.

\placetable{m_zbin}

\pagebreak[0]
\section{The Catalog} \label{catalog}
The results for individual stars are given in Table~\ref{stars}, 
where 
the columns are, respectively: 1) the star name; 
2) the difference between our calculated velocity of the star 
(see \S \ref{radvel}), and the cluster velocity given in the MWGC catalog; 
3) the V magnitude of the star above the horizontal branch; 
4)  the de-reddened \bv\ color of the star determined with \red\ from the MWGC 
catalog;  
5) the weighted mean \ews\ of the
star, calculated according to \S \ref{EW};
6) the external error of the weighted mean \ews; 
7) the probability of cluster membership from proper motion data;
8) the probability of cluster membership from our calculated velocities
(see \S~\ref{vel_mem});
9,10,11) the $f_p, f_r$ values and final weight, $w,$ respectively,  used 
in our fitting technique to calculate the reduced equivalent width, 
\redew, of the  cluster (see \S \ref{reduced_ew} for technique, and
Table~\ref{cluster_res} for \redew\ values);  the weight has been 
scaled in each cluster so that the highest weight star in the cluster
has $w = 1$; 12) the number of spectra analyzed for the star; 13) the
nights that the star was observed, where the first number indicates the
observing run, and the other numbers indicate the  specific night of
observation.  Therefore, 1-1 represents April 13, 1989, 1-2 represents
April 14, 1989 $\ldots$, and 2-1 represents July 13, 1989, $\ldots$
If a star was observed on more that one night for a given run, then it
will have more than one number following the dash. Finally, column
14) indicates when notes 
are found  at the bottom of the table.

The values of \vclush, \vhb, and \red\ for each cluster can be found
in Table~\ref{clusters}.  The references for these values, as well as  
the references for the Table~\ref{stars} star names, photometry, 
and the proper motions (in columns 1,3,4 and 7, respectively) 
can be found in Appendix~\ref{notes}.  Notes for a given cluster, or
stars in the cluster, can also be found in Appendix~\ref{notes}.
In Figure~\ref{ewcmd.grp}, the $\vhb - \vv$ magnitudes are plotted 
against the \ews\ values in the left panel.  The light line
represents the robust line fit (see \S \ref{reduced_ew}) to the cluster
stars, while the bold lines represent the line fits to  three well 
studied clusters spanning a large range in \fe\ and each having low
reddening.
The CMD of the same stars is  plotted in the 
right panel to give an indication of possible non-RGB stars, and
to allow assessment of the photometry, where a large 
scatter is likely indicative of sizable photometric errors, differential
reddening, inclusion of undetected non-members, or some combination thereof.
The RGBs of the same three fiducial clusters 
used in the left panels are plotted in the right panels with bold lines.  
The relative placement, with respect to the fiducial lines, of the 
cluster line fit in the left panels, and the RGB stars in the right 
panels give an indication of the accuracy of the \red\ values listed in 
the MWGC catalog.  The details for these figures are given in the figure 
caption.  The interpretation of the data compiled in this paper will be
presented in subsequent publications.  
  
\placetable{stars}
\placefigure{ewcmd.grp}

\notetoeditor{please place figures 7-24 here, which all fall under the
same caption given for the label ewcmd.grp}

\pagebreak[0]
\acknowledgments

We thank Las Campanas night assistants Angel Guerra and Fernando Peralta 
for their cheerful, effective support, and W.E. Kunkel for setting up the
spectrograph and instructing JEH in its use for run one. The large
quantity of photographic work required to prepare the extensive,
large-scale finding charts essential to the observations was ably
performed by Dave Duncan at DAO.  We thank Gianni Marconi, Hugh 
Harris, Don Terndrup, and Sergio Ortolani for sending us ASCII
copies of their photometry, in some cases before publication.  Thanks
also to Bill Harris for compiling the MWGC catalog, for 
insightful discussions, and for a helpful referee's report.  
GAR would like to thank the NRC for funding during this project.

\pagebreak[0]
\appendix
\section{The Program Clusters} \label{notes}
The reference cluster data (see Table~\ref{clusters}) were taken from
the June 1994 version of the \markcite{harris96} Harris~(1996)
electronic MWGC catalog.  Since this is a dynamic catalog, the
reference sources for the data we adopted are listed below for each
cluster.  The cluster coordinates were taken from
\markcite{djorgovski93} Djorgovski and Meylan~(1993).  The sources for
the \red\ measurements were from \markcite{reed88} Reed et al.~(1988),
\markcite{webbink85} Webbink~(1985), and \markcite{zinn85}
Zinn~(1985).  The catalog documentation states, ``In addition to
the three major sources listed above, measurements of \red\ from the
individual color-magnitude studies (sources given for \vhb ) were
employed whenever they appeared to be well calibrated... The final
adopted reddenings are the straight averages of the given sources (up
to 4 per cluster)''.   The cluster radial velocities were mainly taken
from \markcite{armzinn88} Armandroff and Zinn~(1988),
\markcite{hesser86} Hesser, Shawl and Meyer~(1986),
\markcite{webbink81} Webbink~(1981), and \markcite{zinn84} Zinn and
West~(1984).  Again, the catalog documentation states, ``However,
numerous more recent sources are also available for smaller lists of
objects; in many cases these are based on large samples of stars from
CORAVEL or multi-object echelle spectra with very high precision
($\pm$1 \kms\ or less) and almost totally supersede any previous data.
The adopted $v_r$ for each cluster is the average of the available
measurements, each one weighted inversely as the published
uncertainty.'' These more recent sources are listed below for
individual clusters in the form \vclush: author, author\ldots.  

More generally, for each cluster we list individually the relevant
data sources, as follows.

Horizontal-branch magnitudes (\vhb): these were measured from the
sources, which are cited in the form \vhb: type of data-author, type of
data-author,\ldots, where the type of data is one of: 1) CCD for
photometry done with a charged coupled device, 2) PG for photometry done
with a photographic plate, and 3) RR Lyrae if the determination is
specifically referred to them.

Proper motion data (\pmu): values listed in Table~\ref{stars} are 
cited in the notes here in the form \pmu: author.  

Stellar identifications and photometric zero points:  sources are cited
in the form ID: author[letter], author[letter],\ldots, where the letter
is used to preface the star ID in Table~\ref{stars} and for discussion
of any photometric zero-point adjustments.  Because a 0.1
\vv\ magnitude zero-point offset between photometry used for the HB and
RGB stars would lead to a systematic shift in the calculated \fe\ of
$\sim$0.025~dex for metal-poor clusters, and more for metal-rich
clusters, zero-point offsets larger than $\sim$0.1~mag between
photometric systems were always applied to the RGB stars to bring them
onto the \vhb\ system.  These adjustments are listed below as \vhb\ $-
\vv_{letter}$ = \dots; in more complicated scenarios, they are
explicitly given.  When star-by-star comparisons were performed, the
$\pm$ refers to the standard deviation of the sample.  We also indicate
how the photometry from different sources was combined to give the
\vv\ values listed in Table~\ref{stars}.  If there are no comments for
a cluster, then the photometry was taken unchanged from the single 
source listed.

\paragraph{NGC 104 = M 12 = 47 Tuc -}
\vclush: 
\markcite{arm86} Armandroff and Da Costa (1986),
\markcite{meylan91} Meylan et al. (1991),
\markcite{meylan86} Meylan and Mayor (1986);
\vhb:
\markcite{hesser87} CCD - Hesser et al. (1987);
\pmu:
\markcite{tucholke92a} Tucholke (1992a);
ID:
\markcite{lee77b} Lee (1977b)[L]. 
Hesser et al. (see their Figure~11, and Appendix E) found that their V
photometry agrees with L to within $\pm$ 0.02 mag, so no corrections
were applied.

\pagebreak[0]
\paragraph{NGC 288 -}
\vclush: 
\markcite{peterson86} Peterson et al. (1986),
\markcite{pryor91} Pryor et al. (1991),
\markcite{pryor93} Pryor and Meylan (1993);
\vhb:
\markcite{berg93} CCD - Bergbusch (1993);
ID:
\markcite{alcainoliller80c} Alcaino and Liller (1980c)[A], 
\markcite{olszewski84} Olszewski et al. (1984)[O].
Using 10 stars, the \vv\ photometry of A was brighter than O's by $-0.08
\pm 0.09$ mag, so the straight mean was taken for these stars for both
\vv\ and \bv.  No stars from A or O were in common with the Bergbusch
study, but \markcite{bolte92} Bolte (1992, see his Figure~6) finds good
agreement between his photometry and O. Bergbusch found that his
\vv\ photometry was $\sim0.064$ mag fainter than Bolte, so no
corrections were applied.

\pagebreak[0]
\paragraph{NGC 362 -}
\vclush: 
\markcite{fischer93} Fischer et al. (1993);
\vhb:
\markcite{harris82} PG - Harris (1982);
\pmu:
\markcite{tucholke92b} Tucholke (1992b);
ID:
\markcite{harris82} Harris (1982)[H].

\pagebreak[0]
\paragraph{NGC 1261 -}
\vhb:
\markcite{ferraro93} CCD - Ferraro et al. (1993);
ID:
\markcite{ferraro93} Ferraro et al. (1993)[F],
\markcite{alcaino79b} Alcaino (1979b)[A].
All photometry was taken from from F.

\pagebreak[0]
\paragraph{NGC 2298 -}
\vhb:
\markcite{janes88} CCD - Janes and Heasley (1988);
ID:
\markcite{alcainoliller86a} Alcaino and Liller (1986a)[A].
No stars from A were in common with the Janes and Heasely study, but
\vhb\ in both studies indicates that there is not a significant 
zero-point difference.

\pagebreak[0]
\paragraph{NGC 2808 -}
\vhb:
\markcite{ferraro90} CCD - Ferraro et al. (1990);
ID:
\markcite{harris75} Harris (1975)[H], 
\markcite{harris78} Harris (1978). 
The photometry was taken from Harris (1975), and recalibrated according to 
Harris (1978).  We could not reproduce Ferraro et al.'s comparison to the 
Harris (1978) data, so from our comparison of 14 stars, the \vv\ photometry
of Ferraro et al. was brighter than Harris (1978) by $0.06 \pm 0.04$ mag,
and the \bv\ photometry was redder by $0.09 \pm 0.06$.  Ferraro et al.
photometry was used for stars when available, and otherwise Harris (1978)
data was used with no corrections.

\pagebreak[0]
\paragraph{NGC 3201 -}
\vclush: 
\markcite{cote95} Cot\'e et al. (1995);
\vhb:
\markcite{brewer93} CCD - Brewer et al. (1993);
ID:
\markcite{lee77c} Lee (1977c)[L]. 
Star 2405 was listed in both the PE and PG data of L;  we assumed that
the PG value was correct.  Brewer et al. find systematic differences
with the photometry of L, but this effect was $\lesssim$ 0.1 mag in V,
for the central part of the cluster, and less severe in the outskirts.
Since our stars were selected far from the center, no corrections were
applied.

\pagebreak[0]
\paragraph{NGC 4372 -}
\vhb:
\markcite{alcaino91} CCD - Alcaino et al. (1991);
ID:
\markcite{alcaino74a} Alcaino (1974a)[A]. 
Alcaino et al. find no systematic differences with the A photometry.

\pagebreak[0]
\paragraph{NGC 4590 = M 68 -}
\vclush: 
\markcite{pryor93} Pryor and Meylan (1993);
\vhb:
\markcite{mcclure87} CCD - McClure et al. (1987);
ID:
\markcite{harris75} Harris (1975)[H],
\markcite{alcaino77a} Alcaino (1977a)[A].
Using 12 stars, the \vv\ photometry of H was brighter than A by $-0.03
\pm 0.01$, so no adjustments were made.  McClure et al. (see their
Figure~3) overplotted the CMD of H on their data; no significant
difference in \vhb\ was found.  We obtained two spectra of HI184 with low
$\son \sim20$.  Although we determine this star to be a 94\% probable velocity
member, S93, who obtained higher precision velocity data, showed that this star
is not a velocity member, and thus it was not used in our \redew\ analysis.  

\pagebreak[0]
\paragraph{NGC 4833 -}
\vhb:
\markcite{menzies72} PG - Menzies (1972);
ID:
\markcite{menzies72} Menzies (1972)[M].

\pagebreak[0]
\paragraph{NGC 5286 -}
\vhb:
\markcite{harris76} PG - Harris et al. (1976);
ID:
\markcite{harris76} PG - Harris et al. (1976)[H].

\pagebreak[0]
\paragraph{NGC 5897 -}
\vhb:
\markcite{ferraro92} CCD - Ferraro et al. (1992);
ID:
\markcite{sandage68} Sandage and Katem (1968)[S]. 
Ferraro et al. (see their Figure~3) found that  their \vv photometry
was $\sim0.1$ mag fainter than S, so $\vhb\ - \vv_S = 0.1$.

\pagebreak[0]
\paragraph{NGC 5904 = M 5 -}
\vclush: 
\markcite{olszewski86} Olszewski et al. (1986),
\markcite{peterson86} Peterson et al. (1986),
\markcite{rastorguev91} Rastorguev and Samus (1991);
\vhb:
\markcite{Storm91} CCD photometric RR Lyrae - Storm et al. (1991);
\pmu:
\markcite{cudworth79} Cudworth (1979);
ID:
\markcite{buon81} Buonanno et al. (1981)[B]. 
From the level of \vhb\ in the CMD of B, there does not appear to be a
systematic \vv\ offset from Storm et al.

\pagebreak[0]
\paragraph{NGC 5927 -}
\vhb:
\markcite{saranorris94} CCD - Sarajedini and Norris (1994), 
\markcite{friel91} CCD - Friel and Geisler (1991);
ID:
\markcite{menzies74b} Menzies (1974b)[M]. 
Sarajedini and Norris found that their \vv\ photometry was on average
0.2 mag fainter than M, so $\vhb\ - \vv_M = 0.2$.

\pagebreak[0]
\paragraph{NGC 5986 -}
\vhb:
\markcite{bond94} CCD - Bond et al. (1994);
ID:
\markcite{harris76} Harris et al. (1976)[H]. 
From the level of \vhb\ in the CMD of H, there does not appear to be a 
systematic \vv\ offset from Bond et al.

\pagebreak[0]
\paragraph{NGC 6093 = M 80 -}
\vhb:
\markcite{harris74} PG - Harris and Racine (1974);
ID:
\markcite{harris74} PG - Harris and Racine (1974)[H].

\pagebreak[0]
\paragraph{NGC 6101 -}
\vhb:
\markcite{sara91} CCD - Sarajedini and Da Costa (1991) 
ID:
\markcite{alcaino74b} Alcaino (1974b)[A] 
\markcite{marconi96} Marconi (private communication)[M]
M found that the \vv\ photometry of Sarajedini and Da Costa
was on average 0.08 mag brighter than M, so $\vhb\ - \vv_M = -0.08$.

\pagebreak[0]
\paragraph{NGC 6121 = M 4 -}
\vclush: 
\markcite{clementini94} Clementini et al. (1994),
\markcite{petersonlatham86} Peterson and Latham (1986),
\markcite{peterson86} Peterson et al. (1986),
\markcite{rastorguev91} Rastorguev and Samus (1991);
\vhb:
\markcite{cudworth90} PG - Cudworth and Rees (1990);
\pmu:
\markcite{cudworth90} Cudworth and Rees (1990);
ID:
\markcite{lee77a} Lee (1977a)[L]. 
Photometry was taken from Cudworth and Rees.

\pagebreak[0]
\paragraph{NGC 6144 -}
\vhb:
\markcite{alcaino80} PG - Alcaino (1980);
ID:
\markcite{alcaino80} Alcaino (1980)[A].

\pagebreak[0]
\paragraph{NGC 6171 = M 107 -}
\vclush: 
\markcite{dacosta89} Da Costa and Seitzer (1989),
\markcite{piatek94} Piatek et al. (1994),
\markcite{pryor87} Pryor et al. (1987);
\vhb:
\markcite{cudworth92} PG - Cudworth et al. (1992); 
\pmu:
\markcite{cudworth92} Cudworth et al. (1992); 
ID:
\markcite{sandage64} Sandage and Katem (1964)[S]. 
Photometry taken from Cudworth et al.

\pagebreak[0]
\paragraph{NGC 6218 = M 12 -}
\vclush: 
\markcite{hharris83} Harris et al. (1983),
\markcite{pryor87} Pryor et al. (1987),
\markcite{rastorguev91} Rastorguev and Samus (1991);
\vhb:
\markcite{racine71} PG - Racine (1971);
ID:
\markcite{racine71} Racine (1971)[R - private communication]. 

\pagebreak[0]
\paragraph{NGC 6235 -}
\vhb:
\markcite{liller80a} PG - Liller (1980a);
ID:
\markcite{liller80a} Liller (1980a)[L].

\pagebreak[0]
\paragraph{NGC 6254 = M 10 -}
\vclush: 
\markcite{rastorguev91} Rastorguev and Samus (1991);
\vhb:
\markcite{hurley89} CCD - Hurley et al. (1989);
ID:
\markcite{harris76} Harris et al. (1976)[H]. 
Hurley et al. found that their \vv\ magnitudes were 0.18 mag fainter
than H, so $\vhb\ - \vv_H = 0.18$

\pagebreak[0]
\paragraph{NGC 6266 = M 62 -}
\vhb:
\markcite{caloi87} CCD BV - Caloi et al. (1987);
ID:
\markcite{alcaino78} Alcaino (1978)[A]. 
Using 14 stars the \vv\ photometry of Caloi et al. was fainter 
than A by $0.14\pm 0.04$ mag, so $\vhb - \vv_A = 0.14$ 

\pagebreak[0]
\paragraph{NGC 6273 = M 19 -}
\vhb:
\markcite{harris76} PG - Harris et al. (1976);
ID:
\markcite{harris76} Harris et al. (1976)[H].

\pagebreak[0]
\paragraph{NGC 6304 -}
\vhb:
\markcite{davidge92} CCD - Davidge et al. (1992);
ID:
\markcite{hesser76} Hesser and Hartwick (1976)[H].  Our estimate of the
\vhb\ level in H was 16.15, whereas the estimate from Davidge et al.
was 16.25, so $\vhb\ - \vv_H = 0.1$

\pagebreak[0]
\paragraph{NGC 6352 -}
\vhb:
\markcite{saranorris94} CCD - Sarajedini and Norris (1994);
ID:
\markcite{saranorris94} Sarajedini and Norris (1994)[S],
\markcite{alcaino71} Alcaino (1971)[A],
\markcite{hartwick72} Hartwick and Hesser (1972)[H].
Using 17 stars, the \vv\ photometry of A was fainter than A by only
0.03 $\pm$ 0.09, and the \bv\ photometry of A was redder than H by only
0.006 $\pm$ 0.1.  Using 10 stars, the \vv\ photometry of S was fainter
than H by 0.25 $\pm$ 0.09, and the \bv\ photometry of S was bluer than
H by $-0.21 \pm$ 0.28.  Photometry was taken from S when available.
Otherwise, the mean of H and A was used or, for the cases where H
photometry was not available, A was used.  When photometry was taken
from H and A, 0.25 was added to \vv, and $-0.2$ was added to \bv.

\pagebreak[0]
\paragraph{NGC 6366 -}
\vclush: 
\markcite{dacosta89} Da Costa and Seitzer (1989);
\vhb:
\markcite{hharris93} CCD - Harris (1993);
ID:
\markcite{pike76} Pike (1976)[P]. 
From the level of \vhb\ in the CMD of P, there does not appear to be a
systematic \vv\ offset from Harris.

\pagebreak[0]
\paragraph{NGC 6362 -}
\vclush: 
\markcite{pryor93} Pryor and Meylan (1993);
\vhb:
\markcite{alcainoliller86b} CCD - Alcaino and Liller (1986b);
ID:
\markcite{alcaino72} Alcaino (1972)[A]. 
Using 12 stars, the \vv\ photometry of
Alcaino and Liller was fainter than A by only 0.08 $\pm$ 0.1 mag,
so no correction was applied.

\pagebreak[0]
\paragraph{NGC 6397 -}
\vhb:
\markcite{alcaino87} digitized PG - Alcaino et al. (1987);
ID:
\markcite{alcaino77b} Alcaino (1977b)[A],
\markcite{cannon74} Cannon (1974)[C]
\markcite{alcaino87} Alcaino et al. (1987)[AB].
Photometry taken from AB.

\pagebreak[0]
\paragraph{NGC 6496 -}
\vhb:
\markcite{saranorris94} CCD - Sarajedini and Norris (1994);
ID:
\markcite{arm88} Armandroff(1988)[A],
\markcite{richtler95} Richtler(1995)[R].
Photometry was taken from R. Due to the high quality
photometry in both studies, it is evident that the \vhb\ in the CMD of
Sarajedini and Norris was $\sim$0.05 mag brighter than R, so $\vhb\ -
\vv_H = -0.05$.  Note that the star A68 = R111/112  was listed as one
star in A's photometry, but was resolved into two stars in the
photometry of R.  Stars R111 and R112 have identical magnitudes
(\vv\ = 15.19) and almost identical colours (\bv\ = 1.40 and 1.39 for
R111 and R112, respectively). We assumed it was one star with
\vv\ = 15.19, since we did not resolve these two stars.  This will not
affect our results since the surface gravity and temperature of both
stars should be very similar.

\pagebreak[0]
\paragraph{NGC 6522 -}
\vhb:
\markcite{terndrup94} CCD - Terndrup and Walker (1994, private
communication; photometry without star names),
ID:
\markcite{arp65} Arp (1965)[A]. 
Photometry was taken from Terndrup and Walker, except for
star A15, which did not have a \bv\ mag listed, so the value given in A
was used.  The star A116 was resolved into three fainter stars by Terndrup
and Walker, and was not used in our  analysis.  The photometry for this
star is from A.

\pagebreak[0]
\paragraph{NGC 6535 -}
\vclush: 
\markcite{pryor93} Pryor and Meylan (1993);
\vhb:
\markcite{sara94} CCD - Sarajedini (1994);
ID:
\markcite{liller80} Liller(1980)[L], 
\markcite{sara94} Sarajedini (1994)[S].
Photometry was taken from Sarajedini.

\pagebreak[0]
\paragraph{NGC 6528 -}
\vhb:
\markcite{ortolani92} CCD - Ortolani et al. (1992);
ID:
\markcite{van79} van den Bergh and Younger (1979)[VY], 
\markcite{ortolani92} Ortolani et al. (1992)[O (private communication)].
Photometry was taken from O, except VYII-42, which was taken from VY.
Using four stars fainter than \vv\ = 16.5, the V photometry of O was
fainter than VY's by 0.05 $\pm$ 0.04,  and the \bv\ photometry was redder
by 0.24 $\pm$ 0.03, so a correction of (0,+0.24) was applied to
(\vv,\bv) for the VY photometry of VYII-42.

\pagebreak[0]
\paragraph{NGC 6544 -}
\vhb:
\markcite{alcaino83} PG - Alcaino (1983);
ID:
\markcite{alcaino83} Alcaino (1983)[A].

\pagebreak[0]
\paragraph{NGC 6541 -}
\vhb:
\markcite{alcaino79a} PG - Alcaino (1979a);
ID:
\markcite{alcaino79a} Alcaino (1979a)[A].

\pagebreak[0]
\paragraph{NGC 6553 -}
\vhb:
\markcite{ortolani90} CCD - Ortolani et al. (1990);
ID:
\markcite{hartwick75} Hartwick (1975)[H], 
\markcite{ortolani90} Ortolani et al. (1990)[O (private communication)].
Photometry was taken from O.  Star HII-3 = O140 was not used in
this analysis, since it is part of the RGB turn over as shown in 
Figure~3b of O.  Star HII-59 was not used due to strong TiO bands
(the photometry for this star was taken from H).

\pagebreak[0]
\paragraph{NGC 6624 -}
\vclush: 
\markcite{pryor89} Pryor et al. (1989),
\markcite{pryor93} Pryor and Meylan (1993);
\vhb:
\markcite{saranorris94} CCD - Sarajedini and Norris (1994);
ID:
\markcite{liller78} Liller and Carney (1978)[L], 
\markcite{richtler95} Richtler (1995)[R]. 
Photometry was taken from R for all stars except LIV150 and LI102, for
which L's photometry was used.  Sarajedini and Norris, as well as R,
found systematic differences with L which were correlated
with \vv. Using Figure~2 and 3 of R, a (\vv,\bv) correction of
(+0.2,-0.2) was applied to LIV150, and (+0.1,-0.1) was applied to
L1102. From the level of \vhb\ in the CMD of R, there do not appear to be
systematic differences with the  \vv\ photometry of Sarajedini and
Norris.

\pagebreak[0]
\paragraph{NGC 6626 -}
\vclush: 
\markcite{pryor89} Pryor et al. (1989),
\markcite{pryor93} Pryor and Meylan (1993);
\vhb:
\markcite{rees91} PG - Rees and Cudworth (1991);
ID:
\markcite{alcaino81} Alcaino (1981)[A - written A(ring\#)-(star\#)]. 
Photometry was taken from Rees and Cudworth, except A1-80, and A2-125, 
for which A's photometry was used.  Using 11 stars, the \vv\ photometry
of Rees and Cudworth was fainter than A by 0.1 $\pm$ 0.08 mag, and the \bv\ 
photometry was bluer by 0.03 $\pm$ 0.02 mag, so the (\vv,\bv) corrections 
applied to the A stars was (+0.1,0).

\pagebreak[0]
\paragraph{NGC 6638 -}
\vhb:
\markcite{alcainoliller83} PG - Alcaino and Liller (1983),
\markcite{smith86} spectroscopy of C-type RR Lyrae - Smith and Stryker (1986);
ID:
\markcite{alcainoliller83} Alcaino and Liller (1983)[A].

\pagebreak[0]
\paragraph{NGC 6637 = M 69 -}
\vhb:
\markcite{saranorris94} CCD - Sarajedini and Norris (1994);
ID:
\markcite{hartwick68} Hartwick and Sandage (1968)[H -note that `n' implies the
star is from the inner circle), 
\markcite{richtler95} Richtler (1995)[R],
\markcite{saranorris94} Sarajedini and Norris (1994)[S].
Photometry was taken from either R or S as indicated by the star
names.  Stars that only had photometry from H were not used due to
their photometric uncertainties (see Figure~13 of S, and Figure~4 of
R).  From the level of \vhb\ in the CMD of R, the \vv\ photometry of
Sarajedini and Norris was $\sim$0.1 mag fainter than R, so 0.1 was
added to the V photometry of R.

\pagebreak[0]
\paragraph{NGC 6681 = M70 -}
\vclush: 
\markcite{pryor89} Pryor et al. (1989);
\vhb:
\markcite{mitter94} CCD - Mittermeier et al. (1994);
ID:
\markcite{harris75} Harris (1975)[H]. 
From the level of \vhb\ in the CMD of H, there does not appear to be a
systematic \vv\ offset from Mittermeier et al.

\pagebreak[0]
\paragraph{NGC 6712 -}
\vclush: 
\markcite{grindlay87} Grindlay et al. (1987);
\vhb:
\markcite{cudworth88} PG - Cudworth (1988);
\pmu:
\markcite{cudworth88} Cudworth (1988);
ID:
\markcite{sandage66} Sandage and Smith (1966)[S]. 
Photometry of Cudworth was used for all stars except SB67 and SA34, for
which the photometry of S was used.  Using 8 stars, the \vv\ photometry of
Cudworth was fainter than S by $0.12 \pm 0.04$ mag, and the \bv\ photometry was
bluer by 0.02 $\pm$ 0.06 mag, so the (\vv,\bv) corrections 
applied to the S stars was (+0.12,0).

\pagebreak[0]
\paragraph{NGC 6717 = Pal 9 -}
\vhb:
\markcite{jensen94} CCD - Jensen et al. (1994);
ID:
\markcite{goranskii79} Goranskii (1979)[G]. 
From the level of \vhb\ in the CMD of G, the \vv\ photometry of
Jensen et al. was $\sim$0.7 mag fainter than G, so $\vhb\ - \vv_R =
0.7$.  All the G photometry was measured by eye except stars G16, G23,
G15, and G24, which were measured with an iris-diaphragm photometer.
Star G35 has an uncertain ID.

\pagebreak[0]
\paragraph{NGC 6723 -}
\vhb:
\markcite{fullton93} CCD - Fullton and Carney (1993);
ID:
\markcite{menzies74a} Menzies (1974a)[M]. 
From the level of \vhb\ in the CMD of M, there does not appear to be a
systematic \vv\ offset from Fullton and Carney.  The star MII-7 was not used in
our analysis since it appears to be a blue HB star.

\pagebreak[0]
\paragraph{NGC 6752 -}
\vhb:
\markcite{buon86} PG - Buonanno et al. (1986);
ID:
\markcite{alcaino72} Alcaino (1972)[A], 
\markcite{cannon73} Cannon and Stobie (1973)[C], 
\markcite{buon86} Buonanno et al. (1986)[B].
Using 12 stars, the \vv\ photometry of C was fainter than B by $0.03
\pm 0.05$, so the \vv\ magnitudes were taken to be the straight mean of
C and B, while the \bv\ magnitudes were simply taken from C.  The star
A9 = B2403 was not used as it is most likely a variable, or was
contaminated in B's study, who obtained a 
\vv\ magnitude 0.35 mag fainter than A and C.

\pagebreak[0]
\paragraph{NGC 6809 -}
\vclush: 
\markcite{pryor91} Pryor et al. (1991),
\markcite{pryor93} Pryor and Meylan (1993);
\vhb:
\markcite{lee77d} PG - Lee (1977d);
ID:
\markcite{lee77d} Lee (1977d)[L].

\pagebreak[0]
\paragraph{NGC 6981 -}
\vhb:
\markcite{dickens72a} PG - Dickens (1972a);
ID:
\markcite{dickens72a} Dickens (1972a)[D].

\pagebreak[0]
\paragraph{NGC 7089 -}
\vclush: 
\markcite{arm86} Armandroff and Da Costa (1986);
\vhb:
\markcite{harris75} PG - Harris (1975);
\pmu:
\markcite{cudworth87} Cudworth and Rauscher (1987)
ID:
\markcite{harris75} Harris (1975)[H].
Using 9 stars, the \vv\ photometry of H was fainter than Cudworth and
Rauscher by $0.004 \pm 0.07$ mag, so the \vv\ magnitudes were taken to
be the straight mean of H and Cudworth and Rauscher, while the
\bv\ magnitudes were simply taken from H.

\pagebreak[0]
\paragraph{NGC 7099 -}
\vclush: 
\markcite{pryor93} Pryor and Meylan (1993);
\vhb:
\markcite{bolte87} CCD - Bolte (1987);
ID:
\markcite{dickens72b} Dickens (1972b)[D - PG magnitudes, DP - PE magnitudes],
\markcite{alcainoliller80b} Alcaino and Liller (1980b)[A].
Photometry was taken from D, or DP.  No stars were in common between
Bolte and D, but \markcite{buon88} Buonanno et al. (1988) find no
significant difference between their photometry and D, and since
\vhb\ = 15.1 for both the photometry of Buonanno et al. and Bolte, no
correction was applied to D.  DP17 has an uncertain ID.

\pagebreak[0]
\paragraph{NGC 7492 -}
\vhb:
\markcite{cote91} CCD - Cot\'e et al. (1991);
ID:
\markcite{buon87} Buonanno et al. (1987)[B], 
\markcite{cuffey61} Cuffey (1961)[C].
Photometry was taken from B. The star CR has an uncertain ID.

\pagebreak[0]
\paragraph{Pal 12 -}
\vclush: 
\markcite{arm91} Armandroff and Da Costa (1991);
\vhb:
\markcite{stetson89} CCD - Stetson et al. (1989);
ID:
\markcite{stetson89} Stetson et al. (1989)[S],
\markcite{harris80} Harris and Canterna (1980)[H]. 
Photometry was taken from S, except H4122, which was taken from H.


\clearpage

\begin{table}
\dummytable\label{clusters}
\end{table}

\begin{table}
\dummytable\label{cluster_res}
\end{table}

\begin{table}
\dummytable\label{ew_tech}
\end{table}

\begin{table}
\dummytable\label{bands}
\end{table}

\begin{table}
\dummytable\label{ew_diag}
\end{table}

\begin{table}
\dummytable\label{regression}
\end{table}

\begin{table}
\dummytable\label{m_author}
\end{table}

\begin{table}
\dummytable\label{m_zbin}
\end{table}

\begin{table}
\dummytable\label{stars}
\end{table}

\clearpage



\clearpage

\figcaption[Rutledge.fig1.ps]{Comparison of spectra with different
\son\ values (estimated as explained in the text) illustrate the
degradation as the \son\ values drop below $\sim$15.
\label{soncomp.grp}}

\figcaption[Rutledge.fig2.ps]{A histogram of the distribution of the
\son\ for the 2640 spectra analyzed.  \label{sondist.grp}}

\figcaption[Rutledge.fig3.ps]{The standard deviation of the velocity
measurements for stars observed consecutively at least twice with the
same exposure times are compared to  the mean \son\ values for each set
of multiple observations of the same star.  \label{sonvel.grp}}

\figcaption[Rutledge.fig4.ps]{The difference between the median
velocity calculated for each star and the cluster velocity given in the
Harris (1994) MWGC catalog is plotted for each of the 52 clusters
observed.  The numbers on the abscissa are linked with NGC numbers in
Table~\ref{clusters}.  \label{veldiff.grp}}

\figcaption[Rutledge.fig5.ps]{As in Figure~\ref{sonvel.grp}, but for the 
standard deviation of the \ews\  measurements for stars observed 
consecutively at least twice with the same exposure times.  
\label{sonmofsig.grp}}

\figcaption[Rutledge.fig6.ps]{A comparison of the
\ews\ values in globular cluster stars from various sources.  
The regression coefficients are found in Table~\ref{regression}.
\label{ewcomp.grp}}

\notetoeditor{figures 7 through 24 are all the same format, and should
appear in the paper as a continuation of the figure 7 caption}

\figcaption[Rutledge.fig7-24.ps]{For each cluster, the $V$ 
magnitude above the horizontal branch (\vhb\ values found in Table~\ref{clusters})
is plotted against the  \ews\ in the left panel, and the color-magnitude diagram 
is plotted in the right panel, where the \bv\ values have been de-reddened by the 
cluster \red\ given in Table~\ref{clusters}.  The \em circles \rm represent
stars that have $f_p$ values greater than 0.75, while the 
\em triangles \rm represent the remainder.  The stars that were not 
used to compute the reduced equivalent width, \redew, are labelled with a $\times$ 
symbol.  The slope in the left panel 
is $0.64 \pm 0.02~\mAA\ts{\rm mag}^{-1}$. The error bars in the left panel are 
calculated as $\sigma = (m^2\ts \sigma^2(V) + \sigma^2(\ews))^{0.5}$, where 
$\sigma(V)$ is assumed to be constant for a cluster and is listed in 
Table~\ref{clusters}, and $\sigma(\ews)$ is listed for each star in 
Table~\ref{stars} (plotted if $\sigma > 0.1$~\AA).   The 
light line in the left panel represents the robust line fit for the
stars in the cluster (see \S~\ref{reduced_ew}), while the bold lines 
represent the robust line fit for the three clusters (from left to right) 
NGC\ts 4590 (M\ts 68), NGC\ts 5904 (M\ts 5), and NGC\ts 104 (47\ts Tuc),
all of which have well determined \fe\
and \redew\ values 
and low \red\ values.  In the right panel, the RGB fiducials for each of 
these clusters is plotted (in the same order from left to right); they were taken, 
respectively, from McClure et al. (1987) [bright end from Harris(1975)], Sandquist 
et al. (1996), and Hesser et al. (1987). 
\label{ewcmd.grp}}
\end{document}